\documentclass{article}

\usepackage{arxiv}

\usepackage[utf8]{inputenc} 
\usepackage[T1]{fontenc}    
\usepackage{hyperref}       
\usepackage{url}            
\usepackage{booktabs}       
\usepackage{amsfonts}       
\usepackage{nicefrac}       
\usepackage{microtype}      
\usepackage{cleveref}       
\usepackage{lipsum}         
\usepackage{graphicx}
\usepackage{natbib}
\usepackage{doi}
\usepackage{subcaption}

\hypersetup{
    colorlinks=false,
    pdfborder={0 0 0},
    pdfborderstyle={/S/U/W 0},
}

\usepackage[labelfont=bf]{caption} 
\title{Machine Learning-driven Multiscale MD Workflows: The Mini-MuMMI Experience}

\usepackage{authblk}

\newbox{\orcid}\sbox{\orcid}{\includegraphics[scale=0.06]{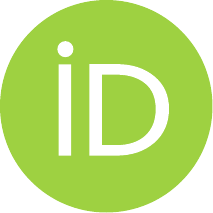}} 
\author[1]{%
	\href{https://orcid.org/0000-0002-7681-3521}{\usebox{\orcid}\hspace{1mm}Lo\"{i}c~Pottier\thanks{Corresponding author: \texttt{pottier1@llnl.gov}}}%
}
\author[2]{Konstantia~Georgouli}%
\author[2]{Timothy~S.~Carpenter}%
\author[2]{Fikret~Aydin}%
\author[2]{Jeremy~O.~B.~Tempkin}%
\author[4]{Dwight~V.~Nissley}%
\author[3]{Frederick~H.~Streitz}%
\author[1]{Thomas~R.~W.~Scogland}%
\author[1]{Peer-Timo~Bremer}
\author[2]{Felice~C.~Lightstone}
\author[2]{%
	\href{https://orcid.org/0000-0002-7613-9143}{\usebox{\orcid}\hspace{1mm}Helgi~I.~Ingólfsson\thanks{Corresponding author: \texttt{ingolfsson1@llnl.gov}}}%
}

\affil[1]{Center for Applied Scientific Computing, Lawrence Livermore National Laboratory, Livermore, 94550, CA, USA}
\affil[2]{Physical and Life Sciences Directorate, Lawrence Livermore National Laboratory, Livermore, 94550, CA, USA}
\affil[3]{Computing Directorate, Lawrence Livermore National Laboratory, Livermore, 94550, CA, USA}
\affil[4]{RAS Initiative, The Cancer Research Technology Program, Frederick National Laboratory, Frederick, 21701, MD, USA}

\renewcommand{\headeright}{}
\renewcommand{\undertitle}{}

\hypersetup{
pdftitle={Machine Learning-driven Multiscale MD Workflows: The Mini-MuMMI Experience},
pdfauthor={Loic Pottier et al.},
pdfkeywords={Molecular dynamics,Scientific workflows,Multiscale,HPC,RAS,RAF,Mini-MuMMI,MuMMI},
}

\begin{document}
\maketitle
\setcounter{footnote}{0}

\begin{abstract}
Computational models have become one of the prevalent methods to model complex phenomena. To accurately model complex interactions, such as detailed biomolecular interactions, scientists often rely on multiscale models comprised of several internal models operating at difference scales, ranging from microscopic to macroscopic length and time scales. Bridging the gap between different time and length scales has historically been challenging but the advent of newer machine learning (ML) approaches has shown promise for tackling that task. Multiscale models require massive amounts of computational power and a powerful workflow management system. Orchestrating ML-driven multiscale studies on parallel systems with thousands of nodes is challenging, the workflow must schedule, allocate and control thousands of simulations operating at different scales. Here, we discuss the massively parallel Multiscale Machine-Learned Modeling Infrastructure (MuMMI), a multiscale workflow management infrastructure, that can orchestrate thousands of molecular dynamics (MD) simulations operating at different timescales, spanning from millisecond to nanosecond. More specifically, we introduce a novel version of MuMMI called "mini-MuMMI". Mini-MuMMI is a curated version of MuMMI designed to run on modest HPC systems or even laptops whereas MuMMI requires larger HPC systems. We demonstrate mini-MuMMI utility by exploring RAS-RAF membrane interactions and discuss the different challenges behind the generalization of multiscale workflows and how mini-MuMMI can be leveraged to target a broader range of applications outside of MD and RAS-RAF interactions.
\end{abstract}


\section{Introduction}
\label{sec:intro}

Large-scale distributed computational platforms (e.g., supercomputers) continue to offer vast amount of raw compute to scientists. The top three supercomputers in the world are now delivering more than one exaflop/s~\cite{RN18}. However, due to their network complexity, intricate memory sub-systems and exotic heterogenous nodes, efficiently utilizing such a massive amount of compute has proven to be challenging~\cite{RN50}. Biologist and chemists have relied on supercomputers capabilities for decades~\cite{RN49,RN20,RN48}. Molecular dynamic (MD) simulations are indeed extremely compute-intensive and require massive amounts of compute to simulate biological systems at a meaningful time and length scales even using different levels of coarse-graining~\cite{RN13,RN14}. Many complex phenomena such as large-scale protein rearrangements are intractable because of the impossibility to harness sufficient compute power to reach the necessary scale and resolution of interest requested by the domain scientists.

There are three ways of approaching that challenge: (i) focus on improving monolithic MD simulation code performance~\cite{RN30,RN29} (e.g., parallelism, use of GPUs, faster solvers) (ii) design a multiscale approach coupling different time and length scales that operates at a coarser grain for given interactions but runs at higher fidelity for more important interactions~\cite{RN17,RN15,RN21} and (iii) extend the multiscale approach to ensemble-based scientific workflows that use ML to automate selection to higher fidelity scales and orchestrate thousands of MD simulation ensembles operating at different scales~\cite{RN1,RN2,RN77}. Even though any progress on the first and second approach ultimately benefits the last one, in this paper we focus on the ensemble-based multiscale approach.

Traditionally, scientists have relied on monolithic computational models operating for a given length and time scale to model complex phenomena. However, more complex interactions like protein-protein interactions and protein structure rearrangement happen at multiple timescales. Atomistic simulations operating at a femtosecond timescale, to capture detailed molecular interactions, are currently computationally intractable to study a phenomenon happening at a millisecond scale. An increasing number of scientific projects that aim to model such interactions rely on multiscale simulations~\cite{RN23,RN19,RN22} and MD is no exception~\cite{RN17,RN21,RN28}. Multiscale modeling offers the possibility to bridge gaps between different time and length scales~\cite{RN21,RN24}. The explosion of computational capabilities has driven the emergence of massive ensembles of simulations operating at different scales orchestrated by complex workflow management systems.

Going hand in hand with multiscale simulation is machine learning (ML)~\cite{RN9}. Multiscale approaches lead to several challenging research questions: how to couple the different scales, how to turn a representation from a given scale to another and how to determine what to sample at a lower scale. Thanks to their ability to extract important features in high-dimensional space, ML models are extremely efficient at multiscale modeling, and several studies have demonstrated the benefits of ML-driven approaches~\cite{RN28,RN7,RN47,RN27}. Among others, the Multiscale Machine-Learned Modeling Infrastructure (MuMMI), has demonstrated the importance of ML-driven ensembles to explore interactions between the RAS proteins and a cell membrane~\cite{RN1,RN2,RN7,RN5}. In addition to ML-related challenges, multiscale approaches pose several significant challenges to HPC systems in terms of resource management. Multiscale ML-driven workflows such as MuMMI are complex pieces of software that rely on hundreds of packages and require expertise in several specific domains (i.e., ML, MD and HPC). All these challenges have dramatically contributed to slowing down the adoption of ML-driven multiscale approaches in the community.

The development of scientific workflows has been key to supporting scientific discoveries~\cite{RN66,RN67}. Ensemble-based approaches are not new~\cite{RN56,RN53,RN54} and have demonstrated their prowess to sample vast spaces and explore large numbers of experimental scenarios for MD~\cite{RN55}, structural biology~\cite{RN24} or material science~\cite{RN57}. Casalino et al.~\cite{RN75} leveraged a supercomputer and an ML-driven multiscale approach to study the mechanisms of COVID-19. Computer scientists have developed many solutions to better support that paradigm, with faster I/O~\cite{RN58} and better in-situ methods~\cite{RN59,RN60,RN61,RN62} so the workflow can analyze data produced by ensembles in near-real time. Several workflow management systems (WMS) have been specifically engineered for ensemble-based runs, such as Ensemble Toolkit~\cite{RN63} or Merlin~\cite{RN64}. However, to the best of our knowledge, no studies have leveraged ML-driven ensemble-based simulations at the scale of MuMMI~\cite{RN1,RN2,RN77,RN7} – running hundreds of thousands of simulations for multiple days. In that sense, MuMMI is unique. Mini-MuMMI offers a unique opportunity for domain and computer scientists to use MuMMI and build upon its foundational effort.

We propose in open-access on GitHub\footnote{\url{https://github.com/mummi-framework}}, mini-MuMMI, a smaller, faster and simpler version of MuMMI. Note that we denote the version of MuMMI used in several publications~\cite{RN1,RN2,RN28,RN7,RN5} as "full", in contrast to the “mini” version of MuMMI discussed in this paper. Mini-MuMMI is a completely functional version of the MuMMI workflow, tailored to be easy to deploy, use and understand so the community can start building on its recognized technology. Mini-MuMMI models the interaction of RAS-RAF, that is KRAS in a complex with the RAS-binding domain (RBD) and cysteine-rich domain (CRD) domains of RAF, with a cell membrane~\cite{RN76}. We chose KRAS because RAS mutations are implicated in roughly a third of all human cancers diagnosed in the US and, KRAS turns out to be the most frequently mutated form of RAS~\cite{RN31,RN32}. Mini-MuMMI comes with a pre-trained ML model based on a slightly modified version of the model showcased in~\cite{RN7} and a set of configuration files for running MD simulations using GROMACS~\cite{RN10}. The system modeled by mini-MuMMI is three to six times smaller than the current model used in large-scale “full” MuMMI campaigns, leading to faster simulations and a workflow that is easier to distribute (i.e., smaller ML model and simulation input files). 

The main goal behind the creation of mini-MuMMI is to foster research in ML-driven multiscale simulations by offering the community a simpler version of MuMMI that can be easily adapted by domain scientists (e.g., used to build experimental campaigns targeting other proteins than RAS, or to use another type of ML modeling approach). The components used in MuMMI publications~\cite{RN1,RN2,RN28,RN7,RN5} can potentially change between studies due to the pursuit of different technical and scientific endeavors over the years i.e., different continuum models, two or three scales. Navigating between the different versions of MuMMI can be tricky, highlighting the need for a stable version that can be used by scientists as a template to learn MuMMI and tune it for their specific needs. The last contribution of this work is more general: we highlight several lessons learned over the years working on MuMMI and, we propose a few key ideas that could improve multiscale workflows support on HPC machines.

\section{General MuMMI concepts}
\label{sec:general}

MuMMI has been used by researchers to explore RAS-RAF interactions and has revealed important lipid-dependent dynamics of RAS signaling proteins~\cite{RN28}. MuMMI is strongly oriented to study RAS-RAF interactions, and even though there have been recent efforts towards the generalization of MuMMI to broaden its scope~\cite{RN2}, MuMMI is still specifically designed for RAS-RAF interactions. Mini-MuMMI is an effort to create a simpler MuMMI comprised of (1) a smaller ML model and (2) a smaller system with fewer atoms simulated. Different flavors of MuMMI have been used for several years in many studies~\cite{RN1,RN2,RN28,RN7,RN5,RN8} making it challenging for scientists to leverage MuMMI architecture. Mini-MuMMI is the first stable version of MuMMI that can be used by researchers as a template to develop their own multiscale modeling workflow. First, we introduce the general concepts and architecture at the heart of mini-MuMMI, then we explain what a MuMMI simulation campaign is. Finally, we highlight the few differences between full (i.e., production) MuMMI and mini-MuMMI. 

\subsection{MuMMI General Objective}

\begin{figure}[h]
	\centering
	\includegraphics[width=1\textwidth]{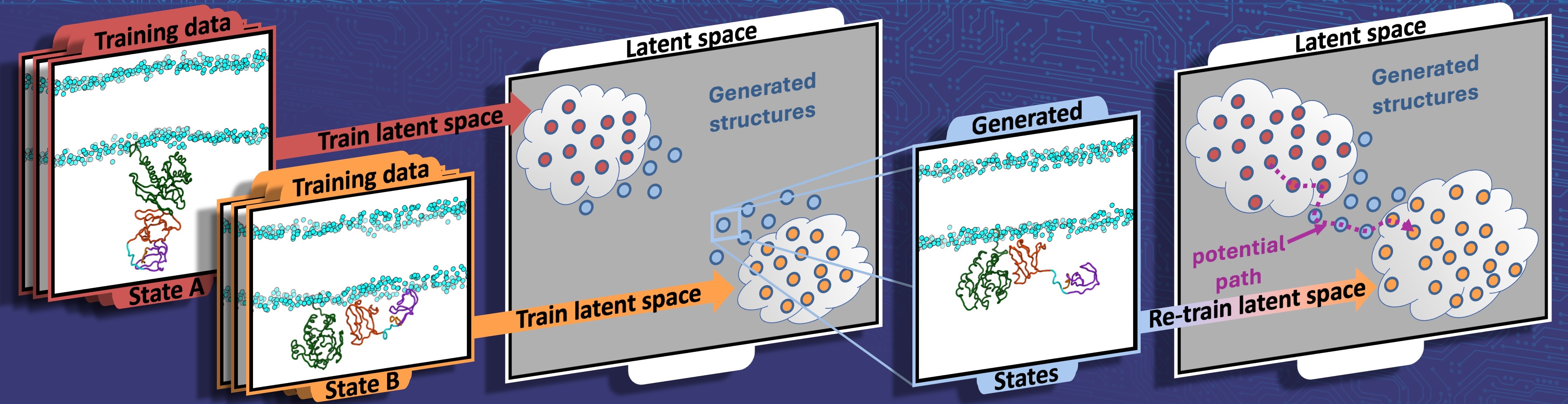}
	\caption{The MuMMI ML-driven multiscale approach. MuMMI aims to find the sequences of conformational states (i.e., paths) connecting two states A and B. MuMMI samples the latent space created by the underlying ML model to find interesting simulations to explore. MuMMI ML model is trained on frames produced by simulations modeling A and B manually set up by domain scientists. The workflow is responsible to coordinate the execution of all these simulations and steer the general sampling towards a potential path between A and B.  In the renderings of the simulation system the lipid headgroups are shown as cyan spheres, RAS is shown in green and the RBD, linker between them, and CRD of BRAF are shown in red, cyan, and purple, respectively.}
	\label{fig:fig1}
\end{figure}

The current configuration of MuMMI~\cite{RN7} tackles a general problem defined as follows: given two protein conformational states, A and B, find a sequence of conformational states connecting A to B (see Figure~\ref{fig:fig1}). To find such sequences of states connecting A to B, a naive approach would be to set up many MD simulations based on state A and let them run long enough to see whether they converge toward B. Unfortunately, this approach is likely to fail in practice: (1) the paths connecting A to B are a priori unknown and (2) even with a known path, such monolithic simulations would have to run for impractical durations (e.g., months or even years) to potentially converge. When trying to model protein interactions, scientists traditionally create one (or a few) monolithic MD simulation(s) designed to run for days, or even months on hundreds of compute nodes. Unfortunately, protein interactions happen at different time and length scales making monolithic simulations very hard to design. Monolithic simulations end up being either long and small or short and large. MD simulations are extremely slow, e.g. all-atomistic simulations with 1.5 million atoms only run at about 14 ns/day on recent GPUs (NVIDIA V100)~\cite{RN2}. Even coarse-grained (CG) simulations~\cite{RN14} which are much faster, barely output 1 $\mu$s/day for a $\sim$140K bead system (CG representation of the same 1.5 million atom system) on one NVIDIA V100~\cite{RN2} while major protein conformational changes typically range from microseconds to milliseconds~\cite{RN69}. Another problem is that the space of potential states between A and B is massive, and many of these intermediate states are likely not leading to a valid path between A and B.

The approach chosen in the current configuration of MuMMI is a ML-driven multiscale approach illustrated in Figure 1 and detailed in~\cite{RN7}. The key behind this approach is the use of an ML model–specifically, an autoencoder setup—capturing the inherent correlations of A and B. That ML model produces a reduced dimensional latent space (LS) which is used to explore the space between A and B. The efficacy of that approach is tied to efficiently sample the LS to extract interesting points that are likely to be on a path. Once a low-dimensional point in the LS has been chosen by the sampling mechanism, the ML model converts (i.e., decodes) that LS point into an actual CG simulation. Then, that prototype of CG simulation is converted into a valid simulation by adding a membrane and a solvent. Finally, the CG simulation is ready to be simulated for a given length by a MD code such as GROMACS.

Note, MuMMI also supports an Ultra-Coarse Grain (UCG) mode where instead of relying on an ML model to sample the LS, MuMMI generates structures by sampling frames produced by multiple concurrently running UCG simulations executed with LAMMPS~\cite{RN68}. UCG is a way to accelerate MD simulations by grouping atoms (or higher resolution CG beads such as Martini CG) into UCG sites based on their collective motions~\cite{RN45,RN44}. This approach is detailed in~\cite{RN8}, in this work we focus on the ML-driven approach detailed in~\cite{RN7}.

MuMMI and mini-MuMMI provide the computational infrastructure to coordinate the execution of tens of thousands of MD simulations and the sampling of a given LS to accelerate biological discovery. MuMMI also iteratively improves its internal sampling process by automatically incorporating feedback from running CG simulations, increasing the likelihood of finding a path~\cite{RN7}.

\subsection{The MuMMI Campaign model}
The first step of a MuMMI campaign consists of setting up the MD simulations that will model states A and B. Then, we manually (i.e., outside of MuMMI) run a few hundred of these simulations to gather enough (i.e., hundreds of thousands) simulation frames to create a meaningful training data set. Then, we train the ML model on the accumulated data. Similar to the previous step, this step is not yet automated and performed manually outside of MuMMI. While building and training the initial ML model is relatively straightforward, the process can be labor-intensive and may demand a deep understanding of the modeled system. The rationale behind the ML models powering MuMMI and mini-MuMMI is the same, however the mini-MuMMI model is trained at a smaller scale with less data. The details of the ML model used in MuMMI are not covered in this publication, interested readers can find details in~\cite{RN7}.

\begin{figure}[h]
	\centering
	\includegraphics[width=0.75\textwidth]{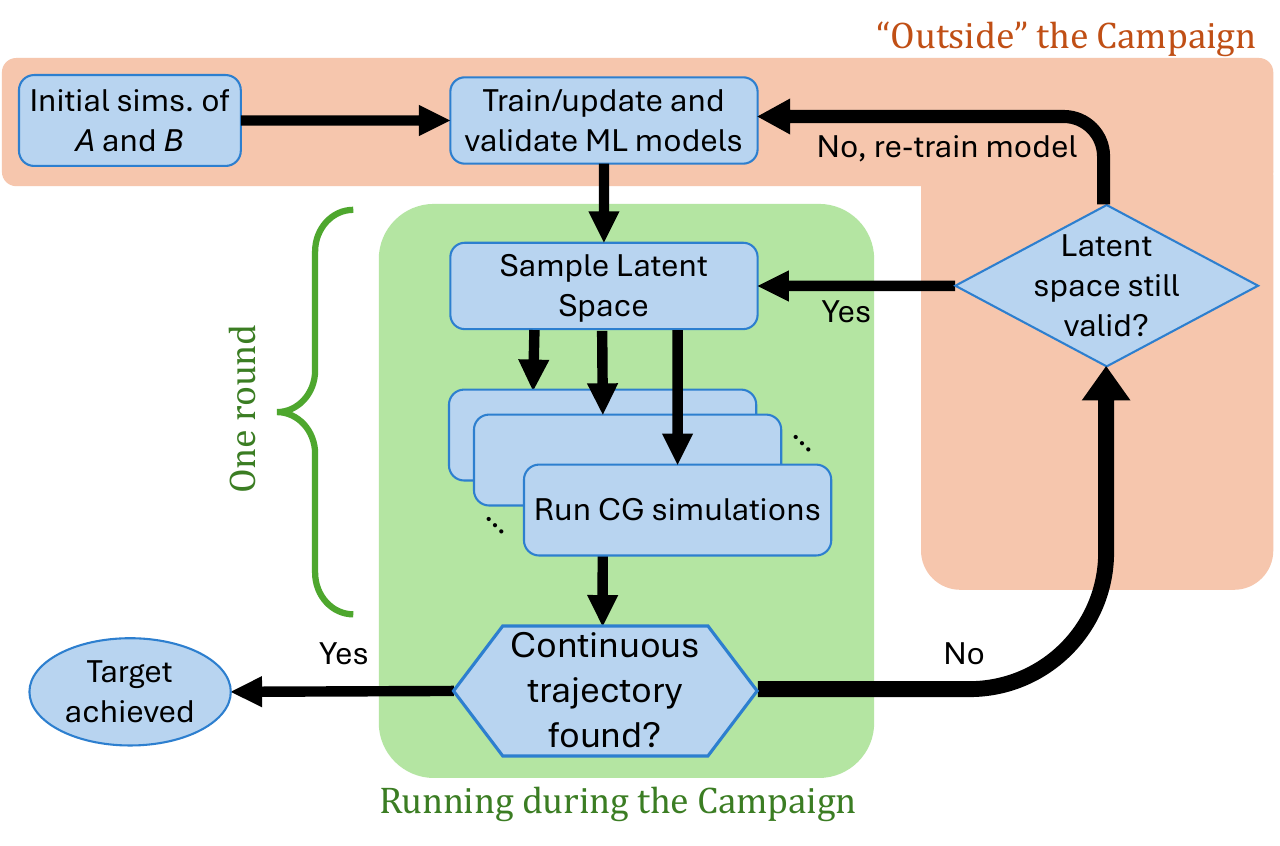}
	\caption{A typical example of a MuMMI campaign. The green part denotes what is managed by the MuMMI infrastructure, the orange part is what domain scientists perform manually, outside of the workflow. The initial simulations for each state are defined and executed manually, as well as the training of the ML model. Once MuMMI has a trained model, it can start exploring the conformational space between states.}
	\label{fig:fig2}
\end{figure}

When all previously described steps have been completed, the part managed by MuMMI can start (green rectangle in Figure~\ref{fig:fig2}). As described in the previous section, MuMMI will start by sampling the LS to find the most promising structures and will then explore these structures by running a given number of CG simulations based on the computing resources available. If MuMMI has sufficiently sampled the LS between A and B, we manually stop the workflow and start the re-training of the ML model, otherwise MuMMI continues until the end of the computing allocation. As mentioned earlier, the ML model is not being retrained on the fly– from time to time MuMMI stops and the ML model is retrained. Such sequence of steps, from sampling the LS to running CG simulations is called a round in MuMMI terminology (see Figure~\ref{fig:fig2}), a round finishes when the ML model is retrained i.e., the LS is updated. A typical campaign consists of 2 to 5 rounds. Ideally, MuMMI would automatically re-train the ML model on-the-fly  based on the amount of new data gathered by MuMMI, but this feature is not supported yet.

\subsection{Mini-MuMMI vs. production MuMMI}
Mini-MuMMI is a pre-configured version of MuMMI, tailored to be easily used by any MD researchers with access to a small HPC cluster or even a laptop. Mini-MuMMI features a fully pre-configured setup for a KRAS-RBDCRD system with $\sim$1280 total lipids ($\sim$600 per leaflet) and about 50K total particles, which is about 3x smaller than the system studied in~\cite{RN7} and 6x smaller than the most recent MuMMI campaign which has not been published yet. Mini-MuMMI comes with a pre-trained autoencoder-based model for that KRAS-RBDCRD system and a set of pre-configured GROMACS input files so MD practitioners can deploy MuMMI out of the box. Mini-MuMMI also comes with various helper scripts to assist users retraining the ML model and analyze the data produced by the simulations. Note that, mini-MuMMI is not different from the full version of MuMMI from a workflow perspective – same jobs, identical data flow and smaller I/Os but identical patterns. However, mini-MuMMI is smaller and simpler with a smaller system, smaller ML model (trained on fewer frames) and no macro-model. The full MuMMI comes with a macro model~\cite{RN43,RN78} used to pre-equilibrate lipids around ML generated protein structures, which for increased simplicity is absent from the mini-MuMMI version.

\section{Mini-MuMMI Workflow}
\label{sec:workflow}

\begin{figure}[h]
	\centering
	\includegraphics[width=0.65\textwidth]{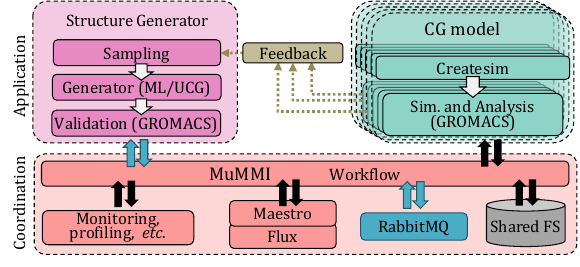}
	\caption{The mini-MuMMI workflow is comprised of two main components: the structure generator (SG) in pink, and the CG simulations (green). These components operate in sequence: starting from the SG producing ML-generated structures to the CG simulations module simulating them. The sampler within the SG consumes feedback files generated by the simulations to iteratively improve its sampling procedure. The workflow manager is responsible for orchestrating the scheduling of simulation ensembles. The workflow communicates with the SG via the message broker RabbitMQ (blue arrows), the rest of communications are done via files (black arrows). White arrows represent control flows between jobs.}
	\label{fig:fig3}
\end{figure}

MuMMI is articulated around two main layers: (1) the application and (2) the orchestration layer. The application layer oversees the MD simulation management, performs in-situ analysis and, in general, any specific analysis or tasks for a given use-case (e.g., RAS-RAF protein interactions). This layer also embeds important decisions like which simulation code (e.g., GROMACS) to use, what analysis to perform or how to store simulation data efficiently i.e., files, database. Scientific domains have usually very different ways of approaching computational simulations, for example biologists and physicists will likely not use the same tools for data management. The application layer also contains several ML components used to sample the LS between the states, and different feedback loops that are important to iteratively improve other components, such as the sampler. The application layer (c.f. Figure~\ref{fig:fig3}) is largely developed by domain experts with the help of computer scientists to extract the best performance from finely tuned MD simulations. In the case of MuMMI, this layer is tailored to study RAS-RAF protein interactions and is developed by experts in MD simulation.

The coordination layer is responsible for orchestrating the simulation ensembles, managing the compute jobs and making sure that all I/Os are done properly, either using files, databases, or a message broker such as RabbitMQ. The orchestration layer is usually written by computer scientists and more generally experts in HPC and scientific workflow management.

\subsection{Application Layer}
\label{sec:app}

The general idea behind mini-MuMMI is that the application layer is fully customizable. The application layer will change based on what problems domain scientists want to tackle. The application layer is comprised of several jobs that are operate in sequence:
\begin{enumerate}
    \item \textit{The structure generator} (SG) is responsible for generating new protein structures for the workflow to explore. The SG is comprised of a sampler, a generator and a validation module. Recall that MuMMI aims to bridge the gap between different protein conformational states and ultimately find a path between these states. The sampler is responsible for selecting novel points to sample between the A and B states by interpolating between points from A and B after being encoded into the LS representation. The sampler uses an on-the-fly feedback mechanism designed to balance exploration vs exploitation based on simulations already witnessed in the sampling process in order to determine which new LS points are worth simulating. Once the sampler has decided which structures are of interest, the generator converts these LS points into actual CG simulations using the decoder part of the model. Finally, the validation module is responsible for validating these newly generated CG simulations by running short energy minimization in vacuum with GROMACS on each of them, if the total energy is within threshold we consider that CG simulation valid.
    \item \textit{Createsims} is a Python script that sets up the CG Martini 3 simulations~\cite{RN41}. The SG generates structures without solvent and membrane, then createsims augments these structures to create complete particle-based simulations with water, salt and 8-component plasma membrane model~\cite{RN74,RN73} using the tool insane~\cite{RN40}. Then, createsims relaxes the membrane and proteins into a more equilibrated state by running energy minimization and equilibrium steps with GROMACS.
    \item \textit{CGAnalysis} is a Python script that is the responsible of running the actual MD simulations based on what createsims generated. CGAnalysis also analyzes the frames produced by the simulation on-the-fly (i.e., in-situ, during the simulation execution). At the beginning, CGAnalysis starts a GROMACS simulation using the simulation setup produced by createsims and, then one or more threads start waiting for frames produced by the simulation. Every X frames produced – where X is a parameter carefully chosen – CGAnalysis runs several analysis procedures and writes feedback files to the filesystem.
\end{enumerate}

\subsection{Coordination Layer}

Mini-MuMMI is managed by a workflow manager (WM) written in Python. The WM is responsible for managing jobs and making sure that the simulations started are the most interesting ones to optimize a given criterion (e.g., in the case of MuMMI, the likelihood to find a path between protein conformational states). The WM is also responsible for requesting new simulations to the SG. The WM and the SG are two independent jobs, and they have to exchange data. To optimize I/Os MuMMI relies on RabbitMQ~\cite{RN38} to communicate between its components. RabbitMQ allows us to scale more easily and has proven to be a valuable tool for inter-job communications (usually performed via inefficient small file I/Os). The management and scheduling of jobs is handled by Maestro~\cite{RN36}, a Python framework that can interface with various HPC schedulers to efficiently schedule thousands of jobs on large-scale platforms. Due to the extreme scale at which MuMMI can operate – tens of thousands of concurrently running jobs – job schedulers like Slurm~\cite{RN35} or IBM Spectrum Load Sharing Facility (LSF)~\cite{RN37} have too many limitations and could not be used. Instead, MuMMI relies on Flux~\cite{RN11}, a modern HPC scheduler developed at Lawrence Livermore National Laboratory, which offers fast job submissions, high flexibility (e.g., hierarchical job submission) and a highly configurable scheduler interface with a Python API. Flux provides many knobs that can be used to fine-tune the scheduler to better suit our needs and improve the job submission rate. In a MD multiscale workflow, it is common for simulations to fail, for example, a non-negligible fraction of createsims fails during a MuMMI campaign due to GROMACS being unable to minimize the energy of some structures i.e., the structures generated by the ML model are not necessarily perfect. Flux allows us to quickly re-submit failed jobs to maximize the utilization of the machine. Furthermore, MuMMI campaigns can span thousands of compute nodes and statistically dozens of nodes will fail~\cite{RN51}. Flux manages failed nodes for us ensuring that jobs running on these nodes are re-scheduled on healthy nodes. Finally, Flux is easy to integrate within MuMMI thanks to its Python-based API.

\begin{figure}[ht]
	\centering
	\includegraphics[width=0.9\textwidth]{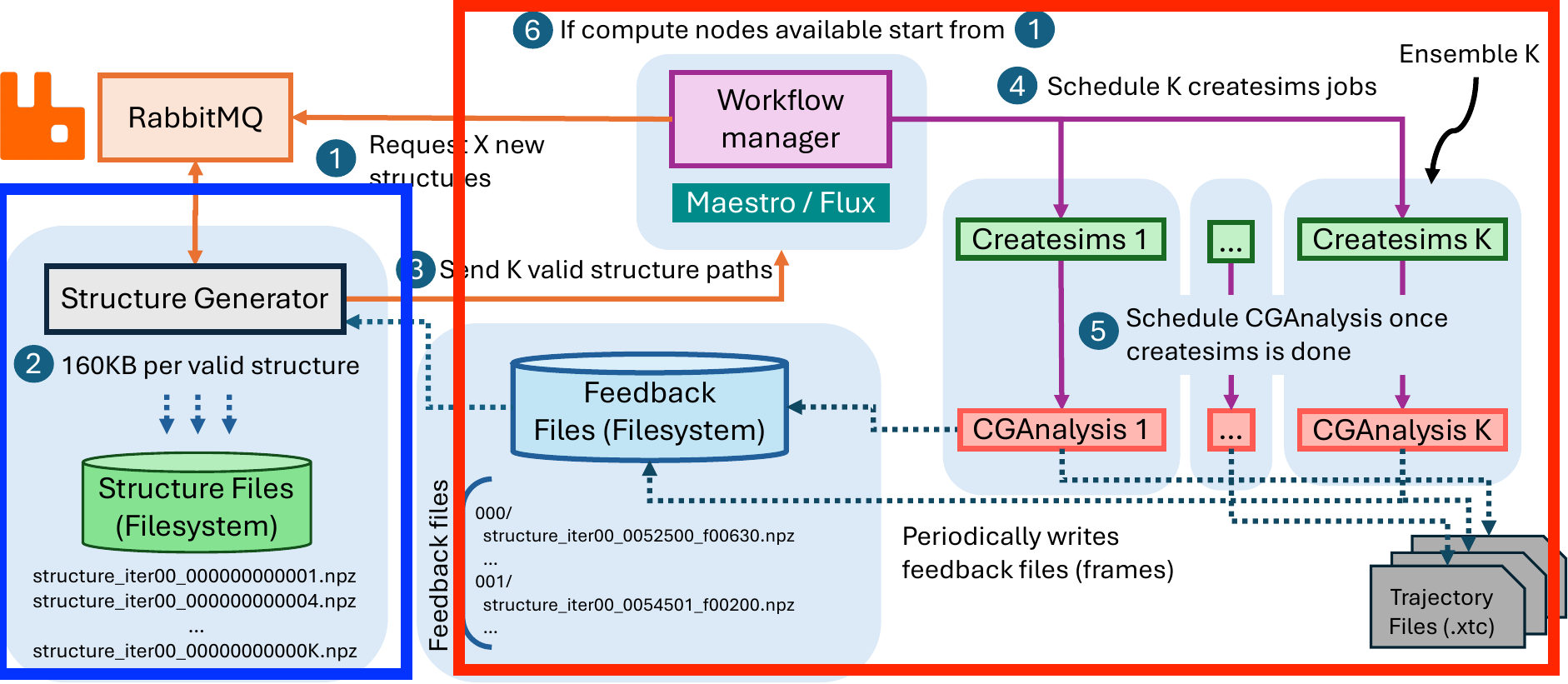}
	\caption{Mini-MuMMI dataflow. The sequence is the following: (1) the WM starts by requesting new structures to explore to Structure Generator (SG), (2) the SG replies with some structures that are deemed worth simulating by the sampler. If there are resources available (GPUs and CPU), the WF schedules a createsims job, once createsims has finished, the WF will schedule the corresponding CGAnalysis. A given createsims and its associated CGAnalysis form an ensemble. Each CGAnalysis will start producing feedback files used to iteratively improve sampling quality.}
	\label{fig:fig4}
\end{figure}

The WM is easily generalizable as it is built upon a highly configurable stack based on various YAML configuration files defining which modes and jobs the workflow should run. Users can easily add new execution modes with their custom job requirements. The Figure~\ref{fig:fig4} describes the dataflow in mini-MuMMI. As previously stated, full and mini-MuMMI have the same logic and dataflow however, the amount of data read/written differs.

\subsection{How to deploy mini-MuMMI}

As previously mentioned, MuMMI relies on hundreds of dependencies, ranging from Python packages to C/C++ codes and, to further darken the picture, MuMMI supports several CPU and GPU architectures, from IBM Power9 and NVIDIA GPUs to AMD CPUs and GPUs. To manage the deployment of MuMMI and automate as many steps as possible, we rely on the HPC package manager Spack~\cite{RN52}. All the Spack environments and custom Spack packages created specifically for MuMMI are available online\footnote{\url{https://github.com/mummi-framework/mummi-spack/}} for everybody to use ensuring complete reproducibility.

\section{Results and Lessons Learned}
\label{sec:results}

\subsection{Simulation Results}

\begin{figure}[ht]
	\centering
    \includegraphics[width=1\textwidth]{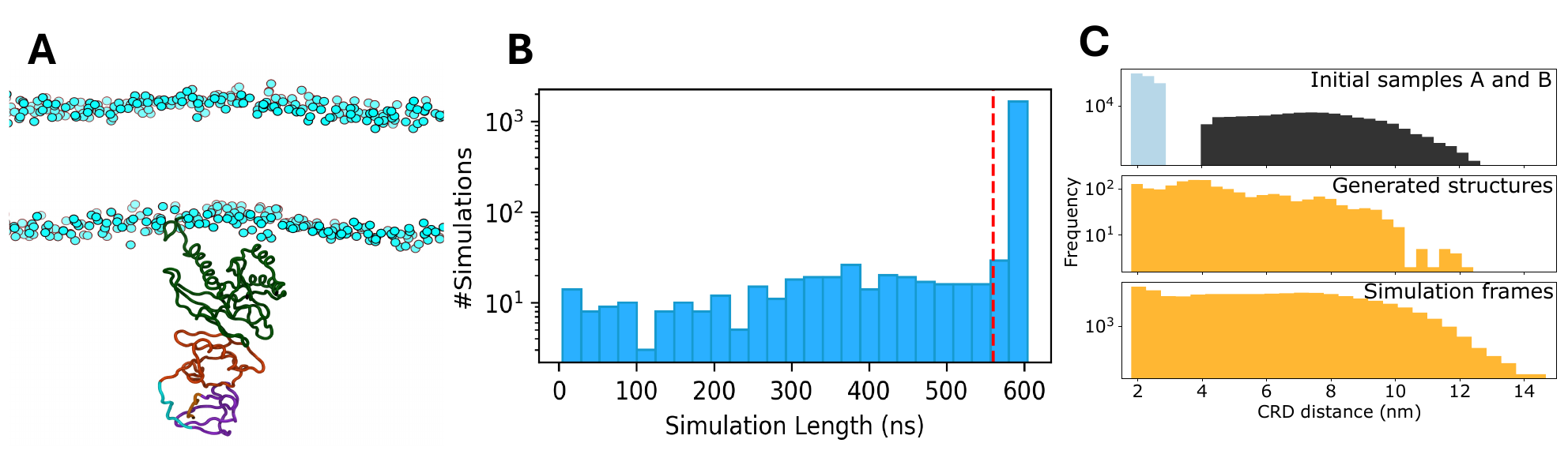}
	\caption{(A) A snapshot of a typical CG simulation showing a zoom in of the membrane associated RAS-RBDCRD protein complex. Water and ions are skipped for clarity, only the phosphate beads of the lipids are shown (cyan beads), and a backbone bead trace is shown for the protein complex. RAS is shown in green and the RBD, linker between them, and CRD of BRAF are shown in red, cyan, and purple, respectively. (B) Distribution of simulation lengths from mini-MuMMI campaign. MuMMI stops CG simulations at 600 ns. The workflow ran 1983 simulations (among which 1632 are longer than 600 ns) for an aggregated simulated time of 1109.4 microseconds. The average simulation length in red is about 575 ns. This campaign ran for 4 hours on 5 nodes and for 36 hours on 64 Frontier nodes. (C) Distribution of the CRD distance for the initial structures representing the two states, ensemble A, CRD membrane distance $\geq$ 4.2 nm; ensemble B, CRD membrane distance $\leq$ 3.2 nm (top panel), the 1983 generated structures using mini-MuMMI between the two conformational states (middle panel), and the frames produced after the CG simulations (bottom panel).}
	\label{fig:fig5}
\end{figure}

We present here results obtained from running a small mini-MuMMI campaign on Frontier, until recently the fastest supercomputer in the world~\cite{RN18}. Frontier is a 1.353 exaflops supercomputer located at Oak Ridge National Laboratory. Frontier has 9,472 compute nodes, each node offers 4 AMD MI250X, one 64-core AMD 3rd Gen EPYC CPU with 512GB of memory and 512GB of GPU memory. Note that, one AMD MI250X contains two Graphics Compute Dies (GCDs). In this paper we consider that one GPU is equivalent to one GCD (e.g., 8 GPUs per node). This demonstration campaign ran for 36 hours on 64 nodes and 4 hours on 5 nodes. Figure~\ref{fig:fig5}A shows the snapshot of one of one the CG simulations generated during that campaign and Figure~\ref{fig:fig5}B depicts the distribution of simulation length achieved by the mini-MuMMI campaign, note that mini-MuMMI has been configured to automatically stop simulations at 600 ns. The mini campaign aggregated about 1.1 milliseconds of CG simulations in 36 hours running on 64 nodes.

Each createsims job uses 1 GPU and 6 CPU cores. Each CGAnalysis is also using 1 GPU and 6 cores. Note that, CGAnalysis requires resources for GROMACS and resources from the in-situ analysis. Usually, one to two cores are dedicated to the Python analysis and the remaining resources are dedicated to the underlying GROMACS simulation. One Frontier node has 64 cores, 8 cores are used by the system and are not usable with specifying specific flags to Slurm (i.e., SLURM core specialization). Thus, one node has 56 usable cores and 8 GPUs, which leads us to 6 cores per job to have 8 simulations per node. Each CGAnalysis job uses 6 cores and 1 GPU, the underlying GROMACS simulation and the analysis share these 6 cores. Table~\ref{tab:table1} presents workflow performance data gathered from several mini-MuMMI and production MuMMI campaigns. Note that, CGAnalysis  jobs in the medium system~\cite{RN7} did not rely on GROMACS but used the MD simulation code ddcMD~\cite{RN34,RN33}.

\begin{table}[h]
    \centering
    \begin{tabular}{lccc}
        \toprule
        & \textbf{Small system} & \textbf{Medium system} & \textbf{Large system\footnotemark} \\
        & \textbf{(mini-MuMMI, this paper)} & \textbf{\cite{RN7}} & \\
        \midrule
        \textbf{System size (\#atoms)} & $\sim$ 50,000 & $\sim$ 132,000 & $\sim$ 300,000 \\
        \midrule
        \textbf{Number of protein beads} & 763 & 751 & 2897 \\
        \midrule
        \textbf{File size per frame} & 0.21 MB & 30 MB & 1.18 MB \\
        \midrule
        \textbf{Snapshot frequency (ns)} & 0.5 & 0.5 & 0.5 \\
        \midrule
        \textbf{Feedback file size} & 8.0 KB & 8.5 KB & 29 KB \\
        \midrule
        \textbf{Createsims time (s)} & 584.2 ($\pm$81.3) & 3602.1 ($\pm$895.6) & 2981.7 ($\pm$369.6) \\
        \midrule
        \textbf{CGAnalysis time (ns/day)} & 2096.7 ($\pm$290.3) & 1179.08 ($\pm$237.3) & 538.1 ($\pm$104.8) \\
        \midrule
        \textbf{ML validation time (s)} & 3.49 ($\pm$0.7) & N/A & 6.43 ($\pm$10.10) \\
        \bottomrule
    \end{tabular}

    \caption{MuMMI workflow data from three different MuMMI campaigns. The snapshot frequency defines the rate at which MD simulations saved frames. Values in seconds/ns are averages, standard deviations in parenthesis. The number of protein beads corresponds to the number of beads (3D position) used in the ML model. The ML validation time is expressed in seconds per structure to validate. The validation time for the medium system is not available due to technical issues. Note that the size per frame for the medium system is larger due to ddcMD storing more information than GROMACS to restart the system (GROMACS stores checkpoints separately).}
    \label{tab:table1}
\end{table}

\footnotetext{Papers analyzing data from this simulation campaign have not yet been publicly released.}

Similar to our previous work~\cite{RN7}, specifically the medium system in Table~\ref{tab:table1}, we denote the two conformational states of interest as A and B. As shown in the top frame of~\ref{fig:fig5}C, the CRD membrane distance, defined as the distance of the CRD domain from the plasma membrane (PM) center, is used to distinguish the structures in the two ensembles~\cite{RN7}. By training an autoencoder using these simulation ensembles, we created a 32-dimensional latent space. This latent space was then used by SG (see Section~\ref{sec:app}) to generate new protein structures between A and B. As a result, a total of 1,983 valid new structures were generated, bridging the two states. The last panel of Figure~\ref{fig:fig5} shows the distribution of the CRD distances for these generated structures. These structures were subsequently used to spawn their corresponding CG simulations with Createsims and CGAnalysis (see Section~\ref{sec:app}). As can be seen in the bottom panel of Figure~\ref{fig:fig5}C, a significant proportion of new simulation frames fill the region between the two simulation ensembles of interest.

This simulation campaign, albeit limited in scale, demonstrates that mini-MuMMI can easily aggregate more than 1 milliseconds of CG simulations and bridge the gap between two conformational states using limited resources.

\subsection{Lessons Learned}

MuMMI has provided computational capabilities to scientists for multiple studies over the years~\cite{RN1,RN2,RN7,RN8}. Thanks to these studies and all the campaigns, we have learned several lessons over the years.

\paragraph{Shared users}
To collaboratively run large multiscale workflows, the support of shared users is required. Most HPC centers do not provide shared users to projects, only one user can run a given job, nobody else can stop the job or check if the job is running correctly, but simulation campaigns like MuMMI are multiple days or even weeks long. Multiple users are needed to ensure that the workflow is running correctly at any given time, as one user cannot be on duty for multiple days. Using a shared user also ensures that file permissions are valid throughout the workflow execution.

\paragraph{Support of network services}
Multiscale workflows generate vast amount of data, e.g., multiple terabytes of data per campaign. But more importantly, a workflow such as MuMMI generates an incredible amount of files, multiple millions, which can lead to I/O quota violations, depending on the HPC center policy. One solution to move from POSIX file-based I/Os is to use an in-memory database such as Redis~\cite{RN12} or RabbitMQ. Unfortunately, deploying network-based services like Redis or RabbitMQ is challenging on HPC platforms, mostly due to HPC centers security requirements. Even though the support of such services by major HPC centers is progressing~\cite{RN71}, the current level of support is still behind cloud-native platforms. To better support future multiscale ensemble of MD simulations, HPC centers should focus on offering better support for container deployment alongside traditional HPC jobs.

\paragraph{Software stack}
The software stack powering MuMMI is large (200+ packages) and complicated, spanning several programming languages. The traditional approaches based on a Python virtual environment or on system modules are not enough for such complex software stack. To deploy MuMMI easily we rely on the Spack package manager, unfortunately Spack is relatively new and not fully supported by many HPC centers. By providing more versatile ways (such as Spack) to manage software dependencies, HPC centers would support workflow development fostering scientific discoveries.

\paragraph{Elastic scheduling}
Coordinating hundreds of thousands of simulations operating at different scales is challenging and requires an HPC scheduler supporting hierarchical scheduling and providing a programmatic API. Thanks to Flux, MuMMI can orchestrate massive size ensembles of simulations ensuring that the workflow maximizes the machine utilization. To efficiently run ML-driven multiscale workflows where different scales exbibit vastly different computational needs (e.g., all-atoms simulations require more resources than coarse-grain simulations) HPC schedulers must better support dynamic scheduling capabilities and provide some elasticity to allocate on-the-fly more resources to a specific part of the workflow. While elasticity is commonly supported in the Cloud realm with Kubernetes~\cite{RN39}, HPC schedulers are behind, although there is some promising work progressing around the convergence between Cloud and HPC~\cite{RN72}. Nevertheless, we believe that multiscale workflows such as MuMMI would greatly benefit from more dynamic scheduling and elastic job allocations.

\section{Conclusion}

We introduced mini-MuMMI, a simplified version of an existing multiscale workflow, specifically tailored for the MD community such that domain scientists can easily build upon our efforts to develop a toolbox for ML driven multiscale simulations. We believe that mini-MuMMI can increase accessibility within the workflow and MD communities and help scientists to better design large-scale ML-driven simulation campaigns using mini-MuMMI as a template. Using Frontier, we ran a small simulation campaign showcasing mini-MuMMI capabilities and we discussed the performance profiles for all components of mini-MuMMI. Computer scientists working on novel scheduling algorithms can struggle to find statistics from realistic workflows (e.g., job duration, I/O sizes) to test their algorithms. By describing the various components of mini-MuMMI and their associated performance and requirements, we provided valuable data for all computer scientists working on workflow resource management and workflow scheduling simulators. MuMMI has been used for more 6 years to ran various simulation campaigns on different systems and HPC centers. Based on these years of experience and the accumulated feedback, we pointed out several lessons we learned along the way and, what we think HPC centers could do to increase the adoption of multiscale workflows. We hope that these recommendations and, the open-sourcing of mini-MuMMI, will push further the adoption of ML-driven multiscale approaches by providing a pre-setup platform that scientists can easily modify and tune to their needs.

Although, we hope that mini-MuMMI contributes to multiscale workflows adoption, there  remain opportunities for improvement. For example, we plan to improve the workflow deployment by packaging mini-MuMMI with containers (i.e., each job of the workflow runs within its dedicated container). Using containers would also allow MuMMI to run easily on Cloud platforms. Additionally, we plan to generalize mini-MuMMI to support multiple MD codes, in addition to GROMACS and ddcMD.

\section*{Acknowledgements}
This work was performed under the auspices of the US Department of Energy (DOE) by Lawrence Livermore National Laboratory under Contract DE-AC52-07NA27344; and under the auspices of the National Cancer Institute (NCI) by Frederick National Laboratory for Cancer Research (FNLCR) under Contract 75N91019D00024. This work has been supported by the NCI-DOE Collaboration established by the US DOE and the NCI of the National Institutes of Health and by Laboratory Directed Research and Development at Lawrence Livermore National Laboratory (24-SI-005). This work was supported by the LLNL Center for Predictive Bioresilience. This research used resources of the Oak Ridge Leadership Computing Facility (OLCF), which is a DOE Office of Science User Facility supported under Contract DE-AC05-00OR22725. For computing time, the authors thank the Advanced Scientific Computing Research Leadership Computing Challenge (ALCC) for time on Frontier and the Livermore Institutional Grand Challenge for time on Lassen. For computing support, the authors thank OLCF and Livermore Computing staff. Release: LLNL-JRNL-2001190.

\bibliographystyle{unsrt}
\bibliography{references}

\begin{thebibliography}{10}

\bibitem{RN18}
J~Dongarra, H~Meuer, and Erich Strohmaier.
\newblock Top 500 supercomputers, 2024.

\bibitem{RN50}
Saman Amarasinghe, Dan Campbell, William Carlson, Andrew Chien, William Dally,
  Elmootazbellah Elnohazy, Mary Hall, Robert Harrison, William Harrod, and
  Kerry Hill.
\newblock Exascale software study: Software challenges in extreme scale
  systems.
\newblock {\em DARPA IPTO, Air Force Research Labs, Tech. Rep}, pages 1--153,
  2009.

\bibitem{RN49}
Rajib Mukherjee, Abhinav Thota, Hideki Fujioka, Thomas~C Bishop, and Shantenu
  Jha.
\newblock Running many molecular dynamics simulations on many supercomputers.
\newblock In {\em Proceedings of the 1st Conference of the Extreme Science and
  Engineering Discovery Environment: Bridging from the eXtreme to the campus
  and beyond}, pages 1--9, 2012.

\bibitem{RN20}
Roland Schulz, Benjamin Lindner, Loukas Petridis, and Jeremy~C Smith.
\newblock Scaling of multimillion-atom biological molecular dynamics simulation
  on a petascale supercomputer.
\newblock {\em Journal of Chemical Theory and Computation}, 5(10):2798--2808,
  2009.

\bibitem{RN48}
Ada Sedova, Russ Davidson, Mathieu Taillefumier, and Wael Elwasif.
\newblock Hpc molecular simulation tries out a new gpu: Experiences on early
  amd test systems for the frontier supercomputer.
\newblock Report, Oak Ridge National Lab.(ORNL), Oak Ridge, TN (United States),
  2022.

\bibitem{RN13}
Siewert~J. Marrink, H.~Jelger Risselada, Serge Yefimov, D.~Peter Tieleman, and
  Alex~H. de~Vries.
\newblock The martini force field: Coarse grained model for biomolecular
  simulations.
\newblock {\em The Journal of Physical Chemistry B}, 111(27):7812--7824, 2007.
\newblock doi: 10.1021/jp071097f.

\bibitem{RN14}
Helgi~I. Ingólfsson, Cesar~A. Lopez, Jaakko~J. Uusitalo, Djurre~H. de~Jong,
  Srinivasa~M. Gopal, Xavier Periole, and Siewert~J. Marrink.
\newblock The power of coarse graining in biomolecular simulations.
\newblock {\em WIREs Computational Molecular Science}, 4(3):225--248, 2014.

\bibitem{RN30}
Frederick~H Streitz, James~N Glosli, Mehul~V Patel, Bor Chan, Robert~K Yates,
  Bronis~R de~Supinski, James Sexton, and John~A Gunnels.
\newblock 100+ tflop solidification simulations on bluegene/l.
\newblock In {\em Proceedings of IEEE/ACM Supercomputing}, volume~5, 2005.

\bibitem{RN29}
J.~N. Glosli, D.~F. Richards, K.~J. Caspersen, R.~E. Rudd, J.~A. Gunnels, and
  F.~H. Streitz.
\newblock Extending stability beyond cpu millennium: a micron-scale atomistic
  simulation of kelvin-helmholtz instability, 2007.

\bibitem{RN17}
Giray Enkavi, Matti Javanainen, Waldemar Kulig, Tomasz Róg, and Ilpo
  Vattulainen.
\newblock Multiscale simulations of biological membranes: The challenge to
  understand biological phenomena in a living substance.
\newblock {\em Chemical Reviews}, 119(9):5607--5774, 2019.
\newblock doi: 10.1021/acs.chemrev.8b00538.

\bibitem{RN15}
Alfons Hoekstra, Bastien Chopard, and Peter Coveney.
\newblock Multiscale modelling and simulation: a position paper.
\newblock {\em Philosophical Transactions of the Royal Society A: Mathematical,
  Physical and Engineering Sciences}, 372(2021):20130377, 2014.

\bibitem{RN21}
Phillip~J. Stansfeld and Mark S.~P. Sansom.
\newblock From coarse grained to atomistic: A serial multiscale approach to
  membrane protein simulations.
\newblock {\em Journal of Chemical Theory and Computation}, 7(4):1157--1166,
  2011.
\newblock doi: 10.1021/ct100569y.

\bibitem{RN1}
Harsh Bhatia, Francesco~Di Natale, Joseph~Y. Moon, Xiaohua Zhang, Joseph~R.
  Chavez, Fikret Aydin, Chris Stanley, Tomas Oppelstrup, Chris Neale,
  Sara~Kokkila Schumacher, Dong~H. Ahn, Stephen Herbein, Timothy~S. Carpenter,
  Sandrasegaram Gnanakaran, Peer-Timo Bremer, James~N. Glosli, Felice~C.
  Lightstone, and Helgi~I. Ingólfsson.
\newblock Generalizable coordination of large multiscale workflows: challenges
  and learnings at scale, 2021.

\bibitem{RN2}
Francesco~Di Natale, Harsh Bhatia, Timothy~S. Carpenter, Chris Neale, Sara
  Kokkila-Schumacher, Tomas Oppelstrup, Liam Stanton, Xiaohua Zhang, Shiv
  Sundram, Thomas R.~W. Scogland, Gautham Dharuman, Michael~P. Surh, Yue Yang,
  Claudia Misale, Lars Schneidenbach, Carlos Costa, Changhoan Kim, Bruce
  D'Amora, Sandrasegaram Gnanakaran, Dwight~V. Nissley, Fred Streitz, Felice~C.
  Lightstone, Peer-Timo Bremer, James~N. Glosli, and Helgi~I. Ingólfsson.
\newblock A massively parallel infrastructure for adaptive multiscale
  simulations: modeling ras initiation pathway for cancer, 2019.

\bibitem{RN77}
Helgi~I. Ingólfsson, Harsh Bhatia, Fikret Aydin, Tomas Oppelstrup, Cesar~A.
  López, Liam~G. Stanton, Timothy~S. Carpenter, Sergio Wong, Francesco
  Di~Natale, Xiaohua Zhang, Joseph~Y. Moon, Christopher~B. Stanley, Joseph~R.
  Chavez, Kien Nguyen, Gautham Dharuman, Violetta Burns, Rebika Shrestha,
  Debanjan Goswami, Gulcin Gulten, Que~N. Van, Arvind Ramanathan, Brian
  Van~Essen, Nicolas~W. Hengartner, Andrew~G. Stephen, Thomas Turbyville,
  Peer-Timo Bremer, S.~Gnanakaran, James~N. Glosli, Felice~C. Lightstone,
  Dwight~V. Nissley, and Frederick~H. Streitz.
\newblock Machine learning-driven multiscale modeling: Bridging the scales with
  a next-generation simulation infrastructure.
\newblock {\em Journal of Chemical Theory and Computation}, 19(9):2658--2675,
  2023.
\newblock doi: 10.1021/acs.jctc.2c01018.

\bibitem{RN23}
Achi Brandt.
\newblock Multiscale scientific computation: Review 2001.
\newblock Multiscale and Multiresolution Methods, pages 3--95. Springer Berlin
  Heidelberg, 2002.

\bibitem{RN19}
Bastien Chopard, Jean-Luc Falcone, Alfons~G. Hoekstra, and Joris Borgdorff.
\newblock A framework for multiscale and multiscience modeling and numerical
  simulations.
\newblock Unconventional Computation, pages 2--8. Springer Berlin Heidelberg,
  2011.

\bibitem{RN22}
Erik Van Der~Giessen, Peter~A Schultz, Nicolas Bertin, Vasily~V Bulatov, Wei
  Cai, Gábor Csányi, Stephen~M Foiles, Marc~GD Geers, Carlos González, and
  Markus Hütter.
\newblock Roadmap on multiscale materials modeling.
\newblock {\em Modelling and Simulation in Materials Science and Engineering},
  28(4):043001, 2020.

\bibitem{RN28}
Helgi~I. Ingólfsson, Chris Neale, Timothy~S. Carpenter, Rebika Shrestha,
  Cesar~A. López, Timothy~H. Tran, Tomas Oppelstrup, Harsh Bhatia, Liam~G.
  Stanton, Xiaohua Zhang, Shiv Sundram, Francesco Di~Natale, Animesh Agarwal,
  Gautham Dharuman, Sara I.~L. Kokkila~Schumacher, Thomas Turbyville, Gulcin
  Gulten, Que~N. Van, Debanjan Goswami, Frantz Jean-Francois, Constance
  Agamasu, De~Chen, Jeevapani~J. Hettige, Timothy Travers, Sumantra Sarkar,
  Michael~P. Surh, Yue Yang, Adam Moody, Shusen Liu, Brian~C. Van~Essen,
  Arthur~F. Voter, Arvind Ramanathan, Nicolas~W. Hengartner, Dhirendra~K.
  Simanshu, Andrew~G. Stephen, Peer-Timo Bremer, S.~Gnanakaran, James~N.
  Glosli, Felice~C. Lightstone, Frank McCormick, Dwight~V. Nissley, and
  Frederick~H. Streitz.
\newblock Machine learning–driven multiscale modeling reveals lipid-dependent
  dynamics of ras signaling proteins.
\newblock {\em Proceedings of the National Academy of Sciences},
  119(1):e2113297119, 2022.

\bibitem{RN24}
Christine Peter and Kurt Kremer.
\newblock Multiscale simulation of soft matter systems – from the atomistic
  to the coarse-grained level and back.
\newblock {\em Soft Matter}, 5(22):4357--4366, 2009.

\bibitem{RN9}
Harsh Bhatia, Fikret Aydin, Timothy~S. Carpenter, Felice~C. Lightstone,
  Peer-Timo Bremer, Helgi~I. Ingólfsson, Dwight~V. Nissley, and Frederick~H.
  Streitz.
\newblock The confluence of machine learning and multiscale simulations.
\newblock {\em Current Opinion in Structural Biology}, 80:102569, 2023.

\bibitem{RN7}
Konstantia Georgouli, Robert~R. Stephany, Jeremy O.~B. Tempkin, Claudio
  Santiago, Fikret Aydin, Mark~A. Heimann, Loïc Pottier, Xiaohua Zhang,
  Timothy~S. Carpenter, Tim Hsu, Dwight~V. Nissley, Frederick~H. Streitz,
  Felice~C. Lightstone, Helgi~I. Ingolfsson, and Peer-Timo Bremer.
\newblock Generating protein structures for pathway discovery using deep
  learning.
\newblock {\em Journal of Chemical Theory and Computation}, 20(20):8795--8806,
  2024.
\newblock doi: 10.1021/acs.jctc.4c00816.

\bibitem{RN47}
F.~Noé, S.~Olsson, J.~Köhler, and H.~Wu.
\newblock Boltzmann generators: Sampling equilibrium states of many-body
  systems with deep learning.
\newblock {\em Science}, 365(6457), 2019.
\newblock Export Date: 31 October 2024; Cited By: 405.

\bibitem{RN27}
Daniel~M. Zuckerman and Lillian~T. Chong.
\newblock Weighted ensemble simulation: Review of methodology, applications,
  and software.
\newblock {\em Annual Review of Biophysics}, 46(Volume 46, 2017):43--57, 2017.

\bibitem{RN5}
Harsh Bhatia, Timothy~S. Carpenter, Helgi~I. Ingólfsson, Gautham Dharuman,
  Piyush Karande, Shusen Liu, Tomas Oppelstrup, Chris Neale, Felice~C.
  Lightstone, Brian Van~Essen, James~N. Glosli, and Peer-Timo Bremer.
\newblock Machine-learning-based dynamic-importance sampling for adaptive
  multiscale simulations.
\newblock {\em Nature Machine Intelligence}, 3(5):401--409, 2021.

\bibitem{RN66}
Aymen Al-Saadi, Dong~H Ahn, Yadu Babuji, Kyle Chard, James Corbett, Mihael
  Hategan, Stephen Herbein, Shantenu Jha, Daniel Laney, and Andre Merzky.
\newblock Exaworks: Workflows for exascale.
\newblock In {\em 2021 IEEE Workshop on Workflows in Support of Large-Scale
  Science (WORKS)}, pages 50--57. IEEE, 2021.

\bibitem{RN67}
Robert Lucas, James Ang, Keren Bergman, Shekhar Borkar, William Carlson, Laura
  Carrington, George Chiu, Robert Colwell, William Dally, and Jack Dongarra.
\newblock Doe advanced scientific computing advisory subcommittee (ascac)
  report: top ten exascale research challenges.
\newblock Report, USDOE Office of Science (SC)(United States), 2014.

\bibitem{RN56}
Yuko Okamoto.
\newblock Generalized-ensemble algorithms: enhanced sampling techniques for
  monte carlo and molecular dynamics simulations.
\newblock {\em Journal of Molecular Graphics and Modelling}, 22(5):425--439,
  2004.

\bibitem{RN53}
Tu~Mai~Anh Do, Loïc Pottier, Rafael Ferreira~da Silva, Silvina Caíno-Lores,
  Michela Taufer, and Ewa Deelman.
\newblock Performance assessment of ensembles of in situ workflows under
  resource constraints.
\newblock {\em Concurrency and Computation: Practice and Experience},
  35(20):e7111, 2023.

\bibitem{RN54}
Maria Böttcher, Alexander Fuchs, Ferenc Leichsenring, Wolfgang Graf, and
  Michael Kaliske.
\newblock Elsa: An efficient, adaptive ensemble learning-based sampling
  approach.
\newblock {\em Advances in Engineering Software}, 154:102974, 2021.

\bibitem{RN55}
Jeffrey Comer, James~C. Phillips, Klaus Schulten, and Christophe Chipot.
\newblock Multiple-replica strategies for free-energy calculations in namd:
  Multiple-walker adaptive biasing force and walker selection rules.
\newblock {\em Journal of Chemical Theory and Computation}, 10(12):5276--5285,
  2014.
\newblock doi: 10.1021/ct500874p.

\bibitem{RN57}
Ayana Ghosh, Bobby~G. Sumpter, Ondrej Dyck, Sergei~V. Kalinin, and Maxim
  Ziatdinov.
\newblock Ensemble learning-iterative training machine learning for uncertainty
  quantification and automated experiment in atom-resolved microscopy.
\newblock {\em npj Computational Materials}, 7(1):100, 2021.

\bibitem{RN75}
Lorenzo Casalino, Abigail~C Dommer, Zied Gaieb, Emilia~P Barros, Terra Sztain,
  Surl-Hee Ahn, Anda Trifan, Alexander Brace, Anthony~T Bogetti, Austin Clyde,
  Heng Ma, Hyungro Lee, Matteo Turilli, Syma Khalid, Lillian~T Chong, Carlos
  Simmerling, David~J Hardy, Julio~DC Maia, James~C Phillips, Thorsten Kurth,
  Abraham~C Stern, Lei Huang, John~D McCalpin, Mahidhar Tatineni, Tom Gibbs,
  John~E Stone, Shantenu Jha, Arvind Ramanathan, and Rommie~E Amaro.
\newblock Ai-driven multiscale simulations illuminate mechanisms of sars-cov-2
  spike dynamics.
\newblock {\em The International Journal of High Performance Computing
  Applications}, 35(5):432--451, 2021.

\bibitem{RN58}
David~A Boyuka, Sriram Lakshminarasimham, Xiaocheng Zou, Zhenhuan Gong, John
  Jenkins, Eric~R Schendel, Norbert Podhorszki, Qing Liu, Scott Klasky, and
  Nagiza~F Samatova.
\newblock Transparent in situ data transformations in adios.
\newblock In {\em 2014 14th IEEE/ACM International Symposium on Cluster, Cloud
  and Grid Computing}, pages 256--266. IEEE, 2014.

\bibitem{RN59}
Ma~Kwan-Liu, Wang Chaoli, Yu~Hongfeng, and Tikhonova Anna.
\newblock In-situ processing and visualization for ultrascale simulations.
\newblock {\em Journal of Physics: Conference Series}, 78(1):012043, 2007.

\bibitem{RN60}
Andrew~C Bauer, Hasan Abbasi, James Ahrens, Hank Childs, Berk Geveci, Scott
  Klasky, Kenneth Moreland, Patrick O'Leary, Venkatram Vishwanath, and Brad
  Whitlock.
\newblock In situ methods, infrastructures, and applications on high
  performance computing platforms.
\newblock In {\em Computer Graphics Forum}, volume~35, pages 577--597. Wiley
  Online Library, 2016.

\bibitem{RN61}
Tu~Mai~Anh Do, Loïc Pottier, Orcun Yildiz, Karan Vahi, Patrycja Krawczuk, Tom
  Peterka, and Ewa Deelman.
\newblock Accelerating scientific workflows on hpc platforms with in situ
  processing.
\newblock In {\em 2022 22nd IEEE International Symposium on Cluster, Cloud and
  Internet Computing (CCGrid)}, pages 1--10. IEEE, 2022.

\bibitem{RN62}
Janine~C Bennett, Hasan Abbasi, Peer-Timo Bremer, Ray Grout, Attila Gyulassy,
  Tong Jin, Scott Klasky, Hemanth Kolla, Manish Parashar, and Valerio Pascucci.
\newblock Combining in-situ and in-transit processing to enable extreme-scale
  scientific analysis.
\newblock In {\em SC'12: Proceedings of the International Conference on High
  Performance Computing, Networking, Storage and Analysis}, pages 1--9. IEEE,
  2012.

\bibitem{RN63}
Vivekanandan Balasubramanian, Antons Treikalis, Ole Weidner, and Shantenu Jha.
\newblock Ensemble toolkit: Scalable and flexible execution of ensembles of
  tasks.
\newblock In {\em 2016 45th International Conference on Parallel Processing
  (ICPP)}, pages 458--463. IEEE, 2016.

\bibitem{RN64}
J~Luc Peterson, Ben Bay, Joe Koning, Peter Robinson, Jessica Semler, Jeremy
  White, Rushil Anirudh, Kevin Athey, Peer-Timo Bremer, and Francesco
  Di~Natale.
\newblock Enabling machine learning-ready hpc ensembles with merlin.
\newblock {\em Future Generation Computer Systems}, 131:255--268, 2022.

\bibitem{RN76}
Timothy~H. Tran, Albert~H. Chan, Lucy~C. Young, Lakshman Bindu, Chris Neale,
  Simon Messing, Srisathiyanarayanan Dharmaiah, Troy Taylor, John-Paul Denson,
  Dominic Esposito, Dwight~V. Nissley, Andrew~G. Stephen, Frank McCormick, and
  Dhirendra~K. Simanshu.
\newblock Kras interaction with raf1 ras-binding domain and cysteine-rich
  domain provides insights into ras-mediated raf activation.
\newblock {\em Nature Communications}, 12(1):1176, 2021.

\bibitem{RN31}
D.~K. Simanshu, D.~V. Nissley, and F.~McCormick.
\newblock Ras proteins and their regulators in human disease.
\newblock {\em Cell}, 170(1):17--33, 2017.
\newblock 1097-4172 Simanshu, Dhirendra K Nissley, Dwight V McCormick, Frank
  HHSN261200800001C/CA/NCI NIH HHS/United States HHSN261200800001E/CA/NCI NIH
  HHS/United States R35 CA197709/CA/NCI NIH HHS/United States Z99
  CA999999/ImNIH/Intramural NIH HHS/United States Journal Article Review United
  States 2017/07/01 Cell. 2017 Jun 29;170(1):17-33. doi:
  10.1016/j.cell.2017.06.009.

\bibitem{RN32}
A.~M. Waters and C.~J. Der.
\newblock Kras: The critical driver and therapeutic target for pancreatic
  cancer.
\newblock {\em Cold Spring Harb Perspect Med}, 8(9), 2018.
\newblock 2157-1422 Waters, Andrew M Der, Channing J R01 CA042978/CA/NCI NIH
  HHS/United States T32 CA009156/CA/NCI NIH HHS/United States P01
  CA203657/CA/NCI NIH HHS/United States P30 CA016086/CA/NCI NIH HHS/United
  States R01 CA175747/CA/NCI NIH HHS/United States U01 CA199235/CA/NCI NIH
  HHS/United States P50 CA196510/CA/NCI NIH HHS/United States Journal Article
  Research Support, N.I.H., Extramural Research Support, Non-U.S. Gov't
  Research Support, U.S. Gov't, Non-P.H.S. Review United States 2017/12/13 Cold
  Spring Harb Perspect Med. 2018 Sep 4;8(9):a031435. doi:
  10.1101/cshperspect.a031435.

\bibitem{RN10}
Szilárd Páll, Mark~James Abraham, Carsten Kutzner, Berk Hess, and Erik
  Lindahl.
\newblock Tackling exascale software challenges in molecular dynamics
  simulations with gromacs.
\newblock In {\em Solving Software Challenges for Exascale: International
  Conference on Exascale Applications and Software, EASC 2014, Stockholm,
  Sweden, April 2-3, 2014, Revised Selected Papers 2}, pages 3--27. Springer,
  2015.

\bibitem{RN8}
Fikret Aydin, Konstantia Georgouli, Loïc Pottier, Timothy~S. Carpenter, Tomas
  Oppelstrup, Xiaohua Zhang, Peer-Timo Bremer, Fred Streitz, Felice~C.
  Lightstone, and Helgi~I. Ingólfsson.
\newblock Enhancing exploration of protein orientational space through
  integration of ultra-coarse-grained models within the mummi.
\newblock {\em Biophysical Journal}, 123(3):237a--238a, 2024.
\newblock doi: 10.1016/j.bpj.2023.11.1502.

\bibitem{RN69}
John~L. Klepeis, Kresten Lindorff-Larsen, Ron~O. Dror, and David~E. Shaw.
\newblock Long-timescale molecular dynamics simulations of protein structure
  and function.
\newblock {\em Current Opinion in Structural Biology}, 19(2):120--127, 2009.

\bibitem{RN68}
Aidan~P Thompson, H~Metin Aktulga, Richard Berger, Dan~S Bolintineanu,
  W~Michael Brown, Paul~S Crozier, Pieter~J In't~Veld, Axel Kohlmeyer, Stan~G
  Moore, and Trung~Dac Nguyen.
\newblock Lammps-a flexible simulation tool for particle-based materials
  modeling at the atomic, meso, and continuum scales.
\newblock {\em Computer Physics Communications}, 271:108171, 2022.

\bibitem{RN45}
E.~Lyman, J.~Pfaendtner, and G.~A. Voth.
\newblock Systematic multiscale parameterization of heterogeneous elastic
  network models of proteins.
\newblock {\em Biophys J}, 95(9):4183--92, 2008.
\newblock 1542-0086 Lyman, Edward Pfaendtner, Jim Voth, Gregory A R01
  GM063796/GM/NIGMS NIH HHS/United States Journal Article Research Support,
  N.I.H., Extramural Research Support, U.S. Gov't, Non-P.H.S. United States
  2008/07/29 Biophys J. 2008 Nov 1;95(9):4183-92. doi:
  10.1529/biophysj.108.139733. Epub 2008 Jul 25.

\bibitem{RN44}
Z.~Zhang, L.~Lu, W.~G. Noid, V.~Krishna, J.~Pfaendtner, and G.~A. Voth.
\newblock A systematic methodology for defining coarse-grained sites in large
  biomolecules.
\newblock {\em Biophys J}, 95(11):5073--83, 2008.
\newblock 1542-0086 Zhang, Zhiyong Lu, Lanyuan Noid, Will G Krishna, Vinod
  Pfaendtner, Jim Voth, Gregory A F32 GM076839/GM/NIGMS NIH HHS/United States 5
  F32 GM076839-02/GM/NIGMS NIH HHS/United States Journal Article Research
  Support, N.I.H., Extramural Research Support, U.S. Gov't, Non-P.H.S. United
  States 2008/09/02 Biophys J. 2008 Dec;95(11):5073-83. doi:
  10.1529/biophysj.108.139626. Epub 2008 Aug 29.

\bibitem{RN43}
T.~Oppelstrup, L.~G. Stanton, J.~O.~B. Tempkin, T.~Ozturk, H.~I. Ingólfsson,
  and T.~S. Carpenter.
\newblock Anisotropic interactions for continuum modeling of protein-membrane
  system.
\newblock {\em The Journal of Chemical Physics}, 2024.

\bibitem{RN78}
L.~G. Stanton, T.~Oppelstrup, T.~S. Carpenter, H.~I. Ingólfsson, M.~P. Surh,
  F.~C. Lightstone, and J.~N. Glosli.
\newblock Dynamic density functional theory of multicomponent cellular
  membranes.
\newblock {\em Physical Review Research}, 5(1):013080, 2023.
\newblock PRRESEARCH.

\bibitem{RN41}
Paulo C.~T. Souza, Riccardo Alessandri, Jonathan Barnoud, Sebastian Thallmair,
  Ignacio Faustino, Fabian Grünewald, Ilias Patmanidis, Haleh Abdizadeh, Bart
  M.~H. Bruininks, Tsjerk~A. Wassenaar, Peter~C. Kroon, Josef Melcr, Vincent
  Nieto, Valentina Corradi, Hanif~M. Khan, Jan Domański, Matti Javanainen,
  Hector Martinez-Seara, Nathalie Reuter, Robert~B. Best, Ilpo Vattulainen,
  Luca Monticelli, Xavier Periole, D.~Peter Tieleman, Alex~H. de~Vries, and
  Siewert~J. Marrink.
\newblock Martini 3: a general purpose force field for coarse-grained molecular
  dynamics.
\newblock {\em Nature Methods}, 18(4):382--388, 2021.

\bibitem{RN74}
Helgi~I. Ingólfsson, Harsh Bhatia, Talia Zeppelin, W.~F.~Drew Bennett,
  Kristy~A. Carpenter, Pin-Chia Hsu, Gautham Dharuman, Peer-Timo Bremer, Birgit
  Schiøtt, Felice~C. Lightstone, and Timothy~S. Carpenter.
\newblock Capturing biologically complex tissue-specific membranes at different
  levels of compositional complexity.
\newblock {\em The Journal of Physical Chemistry B}, 124(36):7819--7829, 2020.
\newblock doi: 10.1021/acs.jpcb.0c03368.

\bibitem{RN73}
T.~N. Ozturk, M.~König, T.~S. Carpenter, K.~B. Pedersen, T.~A. Wassenaar,
  H.~I. Ingólfsson, and S.~J. Marrink.
\newblock Building complex membranes with martini 3.
\newblock {\em Methods Enzymol}, 701:237--285, 2024.
\newblock 1557-7988 Ozturk, Tugba Nur König, Melanie Carpenter, Timothy S
  Pedersen, Kasper B Wassenaar, Tsjerk A Ingólfsson, Helgi I Marrink, Siewert
  J Journal Article United States 2024/07/19 Methods Enzymol. 2024;701:237-285.
  doi: 10.1016/bs.mie.2024.03.010. Epub 2024 Apr 9.

\bibitem{RN40}
Tsjerk~A. Wassenaar, Helgi~I. Ingólfsson, Rainer~A. Böckmann, D.~Peter
  Tieleman, and Siewert~J. Marrink.
\newblock Computational lipidomics with insane: A versatile tool for generating
  custom membranes for molecular simulations.
\newblock {\em Journal of Chemical Theory and Computation}, 11(5):2144--2155,
  2015.
\newblock doi: 10.1021/acs.jctc.5b00209.

\bibitem{RN38}
Jason Williams.
\newblock {\em RabbitMQ in action: distributed messaging for everyone}.
\newblock Simon and Schuster, 2012.

\bibitem{RN36}
Francesco Di~Natale.
\newblock Maestro workflow conductor.
\newblock Report, Lawrence Livermore National Laboratory (LLNL), Livermore, CA
  (United States), 2017.

\bibitem{RN35}
Andy~B Yoo, Morris~A Jette, and Mark Grondona.
\newblock Slurm: Simple linux utility for resource management.
\newblock In {\em Workshop on job scheduling strategies for parallel
  processing}, pages 44--60. Springer, 2003.

\bibitem{RN37}
D~Solt, J~Hursey, A~Lauria, D~Guo, and X~Guo.
\newblock Scalable, fault-tolerant job step management for high performance
  systems.
\newblock 2019.

\bibitem{RN11}
Dong~H Ahn, Ned Bass, Albert Chu, Jim Garlick, Mark Grondona, Stephen Herbein,
  Helgi~I Ingólfsson, Joseph Koning, Tapasya Patki, and Thomas~RW Scogland.
\newblock Flux: Overcoming scheduling challenges for exascale workflows.
\newblock {\em Future Generation Computer Systems}, 110:202--213, 2020.

\bibitem{RN51}
Daniel Dauwe, Sudeep Pasricha, Anthony~A Maciejewski, and Howard~Jay Siegel.
\newblock Resilience-aware resource management for exascale computing systems.
\newblock {\em IEEE Transactions on Sustainable Computing}, 3(4):332--345,
  2018.

\bibitem{RN52}
Todd Gamblin, Matthew LeGendre, Michael~R Collette, Gregory~L Lee, Adam Moody,
  Bronis~R De~Supinski, and Scott Futral.
\newblock The spack package manager: bringing order to hpc software chaos.
\newblock In {\em Proceedings of the International Conference for High
  Performance Computing, Networking, Storage and Analysis}, pages 1--12, 2015.

\bibitem{RN34}
Frederick~H. Streitz, James~N. Glosli, and Mehul~V. Patel.
\newblock Beyond finite-size scaling in solidification simulations.
\newblock {\em Physical Review Letters}, 96(22):225701, 2006.
\newblock PRL.

\bibitem{RN33}
Xiaohua Zhang, Shiv Sundram, Tomas Oppelstrup, Sara I.~L. Kokkila-Schumacher,
  Timothy~S. Carpenter, Helgi~I. Ingólfsson, Frederick~H. Streitz, Felice~C.
  Lightstone, and James~N. Glosli.
\newblock ddcmd: A fully gpu-accelerated molecular dynamics program for the
  martini force field.
\newblock {\em The Journal of Chemical Physics}, 153(4), 2020.

\bibitem{RN12}
J~Carlson.
\newblock {\em Redis in Action}.
\newblock Manning, 2013.

\bibitem{RN71}
Andrew~J Younge, Anthony~Michael Agelastos, and Kevin Pedretti.
\newblock Container utilization at doe compute facilities.
\newblock Report, Sandia National Lab.(SNL-NM), Albuquerque, NM (United
  States), 2020.

\bibitem{RN39}
Brendan Burns, Joe Beda, Kelsey Hightower, and Lachlan Evenson.
\newblock {\em Kubernetes: up and running}.
\newblock " O'Reilly Media, Inc.", 2022.

\bibitem{RN72}
Daniel~J Milroy, Claudia Misale, Giorgis Georgakoudis, Tonia Elengikal, Abhik
  Sarkar, Maurizio Drocco, Tapasya Patki, Jae-Seung Yeom, Carlos Eduardo~Arango
  Gutierrez, and Dong~H Ahn.
\newblock One step closer to converged computing: Achieving scalability with
  cloud-native hpc.
\newblock In {\em 2022 IEEE/ACM 4th International Workshop on Containers and
  New Orchestration Paradigms for Isolated Environments in HPC (CANOPIE-HPC)},
  pages 57--70. IEEE, 2022.

\end{thebibliography}

\end{document}